%% file: p2310tav.tex
\documentclass[12pt]{iopart}
 \usepackage[]{latexsym}
 \usepackage[]{amsfonts}
 \usepackage[]{eepic}
 \usepackage[]{graphics}
\begin{document}
\title[GRECP/MRD-CI calculation of PbH]
{GRECP/5e-MRD-CI calculation of the electronic structure of PbH.}
\author{T.\ A.\ Isaev\footnote[3]{
  E-mail for correspondence: TimIsaev@pnpi.spb.ru;
  http://www.qchem.pnpi.spb.ru},
 N.\ S.\ Mosyagin, A.\ V.\ Titov}
\address{Petersburg Nuclear Physics Institute,
         Gatchina, St.-Petersburg district 188300, RUSSIA}
\author{A.\ B.\ Alekseyev, R.\ J.\ Buenker}
\address{Theoretische Chemie, Bergische Universit{\"a}t Wuppertal,
         Gau{\ss}stra{\ss}e 20, D-42097 Wuppertal, GERMANY}
\date{\today}

\begin{abstract}
 The correlation calculation of the electronic structure of PbH is carried out
 with the Generalized Relativistic Effective Core Potential (GRECP) and
 MultiReference single- and Double-excitation Configuration Interaction
 (MRD-CI) methods. The 22-electron GRECP for Pb is used and the outer core
 $5s$, $5p$ and $5d$ pseudospinors are frozen using the level-shift technique,
 so only five external electrons of PbH are correlated.  A new configuration
 selection scheme with respect to the relativistic multireference states is
 employed in the framework of the MRD-CI method. The [6,4,3,2] correlation
 spin-orbit basis set is optimized in the coupled cluster calculations on the
 Pb atom using a recently proposed procedure, in which functions in the
 spin-orbital basis set are generated from calculations of different ionic
 states of the Pb atom and those functions are considered optimal which
 provide the stationary point for some energy functional. Spectroscopic
 constants for the two lowest-lying electronic states of PbH ($^2\Pi_{1/2},
 ^2\Pi_{3/2}$) are found to be in good agreement with the experimental data.

\end{abstract} 

\maketitle



\section{Introduction}
 \label{Intro}

The Relativistic Effective Core Potential (RECP) method has been widely used
in recent years in the correlation calculations of molecules with heavy and
super-heavy atoms (see e.g.\ \cite{DiLabio}). This method allows one to treat
explicitly only those electrons which play the most important role in the
making of chemical bonds, while interactions with the core electrons and
relativistic effects are described with the help of some effective
one-electron operator (usually semi-local) in the Hamiltonian. Depending on
the desired or reachable level of accuracy, one can choose an effective
Hamiltonian which provides the minimal computational effort for the given
accuracy.  Accordingly, the accuracy of the method used for treating
correlations and that of the RECP method should be consistent. From that point
of view, the PbH molecule is of considerable interest as a testing system.  In
calculations on the Pb atom~\cite{Isaev1}, a comparable level of accuracy for
the Generalized RECP (GRECP) and correlation calculation methods was observed.
PbH is a weakly bound system so the accuracy of the correlation calculation on
the molecule can be expected to be close to the accuracy of the correlation
calculation on the Pb atom.

The GRECP method \cite{GRECPTheory} developed in PNPI allows one to attain a
level of accuracy which is better than 200 cm$^{-1}$ for transition energies
in correlation calculations of different heavy atoms and their ions
\cite{Isaev1,MosHg}. Our recent calculations \cite{TlH} also show that the
MRD-CI method with an improved selection scheme and properly chosen basis set
enables one to reach a level of accuracy for molecular calculations involving
heavy atoms which is comparable to the accuracy of the GRECP calculations for
transition energies between electronic terms in those atoms.

It can also be noted that from valence CI calculations and subsequent
calculations by the Relativistic Coupled Cluster method with Single and Double
excitations (RCC-SD) with the same number of the correlated electrons, one can
easily calculate contributions to the total energy from higher-order cluster
amplitudes first of all in order to account for nondynamic correlation effects
in the valence region.  That information can be used in subsequent
calculations by the RCC-SD method with an increased number of correlated
electrons to give a significant improvement in the accuracy for the
description of correlation structure of the system of interest \cite{Isaev1}.

\section{Methods}

A detailed description of the methods used in the present calculations can be
found elsewhere \cite{GRECPTheory,TlH,MRDCI}. Here we give only brief
description and specifics of each method.

The GRECP operator contains non-local terms with the projectors on outer-core
pseudospinors, together with the radially-local terms common for other RECPs.
The addition of non-local terms is connected with a distinction between the
effective potentials for outer-core and valence electrons. It should be noted
that one-electron integral calculations with GRECPs are only slightly more
time-consuming than with ordinary radially-local RECPs, but this time is
negligible in comparison with the time needed for two-electron integral
calculations.

In the MRD-CI method \cite{MRDCI} only those Spin (and Space Symmetry) Adapted
Functions (SAFs) are included in the CI calculation which provide the values
of the second-order perturbation theory
 corrections to the energy

\begin{equation}
   \label{T}
   \Delta E_I = \frac{|<\Phi_I|H|\Psi_0>|^2}{E_I-E_0}
\end{equation}
which are greater than some threshold $T_0$. In the above equation, $\Psi_0$ is
some multireference function and \{$\Phi_I$\} are the trial SAFs.  As, in
particular, was shown in \cite{TlH} for systems with large Spin-Orbit (SO)
interaction, it is preferable to choose $\Psi_0$ as an eigenvector of some
Hamiltonian including SO terms. The perturbation corrections \cite{MRDCI} for
the $T{=}0$ threshold to the total state energies are applied after the final
SO-CI calculations and Davidson corrections \cite{T0corr} are calculated for
each term.

\section{Basis set}

To generate and optimize basis set for heavy atoms we mainly followed the
procedure described in detail in \cite{Isaev1,MosHg}. The main features of
that procedure are:

\begin{enumerate}

\item[1.] A number of the trial one-electron basis functions is
generated in Dirac-Fock or Hartree-Fock calculations of different ionic states
of the atom of interest with the frozen core orbitals.  In our case,
two-component Hartree-Fock calculations with the spin-averaged part of the
GRECP (Averaged GRECP or AGREP) were carried out. As a result, one-electron
functions with the same $n$ and $l$ quantum numbers become degenerate. Core
orbitals are obtained from the AGREP/SCF calculations of the Pb$^{2+}$ ion
(5s,p,d and 6s orbitals in the case of a 22-electron GRECP). Then SCF
calulations are performed with a different number of electrons in the
frozen-core orbitals and an electron occupying some valence orbital (e.g.\ 6p). 

\item[2.] The correlation calculation is performed with each generated
function included in the basis set.  We carried out RCC-SD calculations of
five lowest-lying states of the Pb atom with the 6s$^2$ 6p$^2$ leading
configuration (see \cite{Isaev1} for more computational details).  In these
calculations the $j$-dependent part of the GRECP (Effective Generalized
Spin-Orbit Potential or EGSOP) is included in the effective Hamiltonian
together with AGREP.  Thus, effects caused by the SO interaction are taken
into account on the basis set of many-electron spin-orbital functions.  It is
clear that for the adequate description of the correlation effects the number
of stored nonidentical two-electron integrals on the spin-orbital basis set is
substantially smaller than that on the spinor basis set.  Therefore,
calculations on the spin-orbital basis sets can be more efficient if
correlation is more important than spin-orbit interaction as it usually takes
place for valence electrons. 

\item[3.] The value of some energy functional is calculated using the
correlation energies obtained in the previous calculations. If the change of
this functional (as compared to the value of the functional obtained in the
previous step) for some function is maximal in comparison with the cases of
using other generated functions and exceeds a given threshold, that function
is added to the correlation basis set. Otherwise, we neglect the function and
the procedure starts from step 1 for an orbital with other $n$ and $l$ quantum
numbers.

\end{enumerate}

\section{Results}

We have carried out three series of MRD-CI calculations for the four
 states going to the two lowest dissociation limits
 of the PbH molecule.  Five external electrons are correlated and the
 $5s_{1/2}$, $5p_{1/2,3/2}$ and $5d_{3/2,5/2}$ pseudospinors are frozen
from the two-component GRECP/SCF calculation of the ground state of the
Pb$^{2+}$ ion (configuration $6s_{1/2}^2$).  At the first run, a few of the
lowest-energy configuration functions in irreducible representations of the
nonrelativistic symmetry group were taken for each electronic term at each
point of the potential curves as the initial approximation for $\Psi_0$ (i.e.\
reference configurations).  For the next two runs, those configurations are
chosen as the reference configurations which give the largest absolute values
of CI coefficients, $C_I$, at the previous run.  The number of the reference
configurations is selected using ``CNFSORT'' code \cite{CNFSORT} such that
their total contribution to the wavefunction ($C^2_{ref} \equiv \sum_{I \in
ref} C^2_I $) is approximately equal to 96\% at the second run and to 98\% at
the third run for each of the considered states and internuclear distances of
the PbH molecule (for more details, see paper by Mosyagin et al.\ from the
same volume and paper \cite{TlH}). 

As ``MRD-CI'' program package only allows the use of symmetry group $D_{2h}$
and lower orders, all calculations have been performed in the relativistic
double group symmetry $C_{2v}$. Molecular spin-orbitals are obtained from SCF
calculations performed by program module ``SCF'' in ``MOLCAS'' package
\cite{MOLCAS}. The correlation spin-orbital basis set for Pb [6,4,3,2]
generated in the way described above and basis set from \cite{DunningJCP}
reduced to [3,2,1] for H are used. We estimate Basis Set Superposition Error
(BSSE) by calculating the Pb atom in the molecular basis set, i.e.\ with
``ghost'' H atom at different distances.  In our calculations the BSSE does
not exceed 70 cm$^{-1}$. The following formula is used for the $T{=}0$ energy
correction:

\begin{equation}
 E_{T{=}0}\ \simeq\ E_{T{=}T_k} -\ \lambda{\sum_{\Delta E_I < T_k}}\Delta E_I\ ,
\end{equation}
where $E_{T{=}T_k}$ and $E_{T{=}0}$ are the total state energies calculated for
the $T_k$ and zero thresholds, $\Delta E_I$ is the energy lowering for the
unselected SAFs estimated by equation~(\ref{T}). The $\lambda$ constant is
determined from calculations with two different thresholds, $T_1$ and $T_2$:

\begin{equation}
   E_{T{=}T_1}\ -\ {\lambda}{\sum_{\Delta E_I<T_1}}\Delta E_I\quad   =\quad
   E_{T{=}T_2}\ -\ {\lambda}{\sum_{\Delta E_I<T_2}}\Delta E_I\ .
\end{equation}
When applying the perturbation energy correction, the value of $\lambda$ never
exceeds 3 with the thresholds from 0.5$\mu H$ to 0.03$\mu H$.  Potential
curves for the four lowest-lying electronic states are presented on Figure~1.
Molecular spectroscopic constants were calculated by the Dunham method in the
Born-Oppenheimer approximation using the DUNHAM-SPECTR code~\cite{Mitin}.

Spectroscopic constants for the two lowest-lying molecular electronic terms
are given in
 Table~1.
The next two electronic levels can not be characterized in the same way, as
can be seen from Figure~1.

 \ack

This work was supported by DFG/RFBR grant No 96--03--00069.  T~I, N~M and A~T
also thank RFBR grants No 01--03--06334, 01--03--06335, and 99--03--33249.
Essential part of the present calculations was carried out in the computer
center of the Bergische Universit{\"a}t Wuppertal.

~\\
{\bf References}
~\\

\newpage
\begin{table}
\caption {Spectroscopical constants ($R_e$~---~bond length, 
$D_0$~---~dissociation energy, $\omega_e$~---~vibrational constant,
$T_e$~---~transition energy) for the ground and first excited states of PbH.}
\begin{tabular}{lll}
 \label{SpectrT}
                  &           &               \\
\cline{1-3}
                  & Calculated& Experimental  \\
 Term             & values    & values\cite{Herzberg} \\
\cline{1-3}
                  &                           & \\
  $ ^2\Pi_{1/2} $ & $R_e = 1.871 $ \AA        &  $R_e = 1.838$ \AA \\
                  & $\omega_e = 1686 cm^{-1}$ & $\omega_e = 1564 cm^{-1}$ \\
                  & $D_0 =1.44 eV$            &  $D_0 \leq 1.59 eV$         \\
                  &                           & \\
  $ ^2\Pi_{3/2} $ & $R_e = 1.855$ \AA         & \\
                  & $\omega_e = 1727 cm^{-1}$ & \\
                  & $ T_e=6427 cm^{-1}$ & $T_e \sim 6900 cm^{-1}$ $^{\rm a)}$ \\
\cline{1-3}
                  &  		              & \\

\end{tabular}

\noindent $^{\rm a)}$ Unpublished data by Fink et al. \\
\end{table}

\input PbH.tex

\end{document}

%% file: PbH.tex
\begingroup%
  \makeatletter%
  \newcommand{\GNUPLOTspecial}{%
    \@sanitize\catcode`\%=14\relax\special}%
  \setlength{\unitlength}{0.1bp}%
\begin{picture}(3600,2260)(300,100)%
\includegraphics{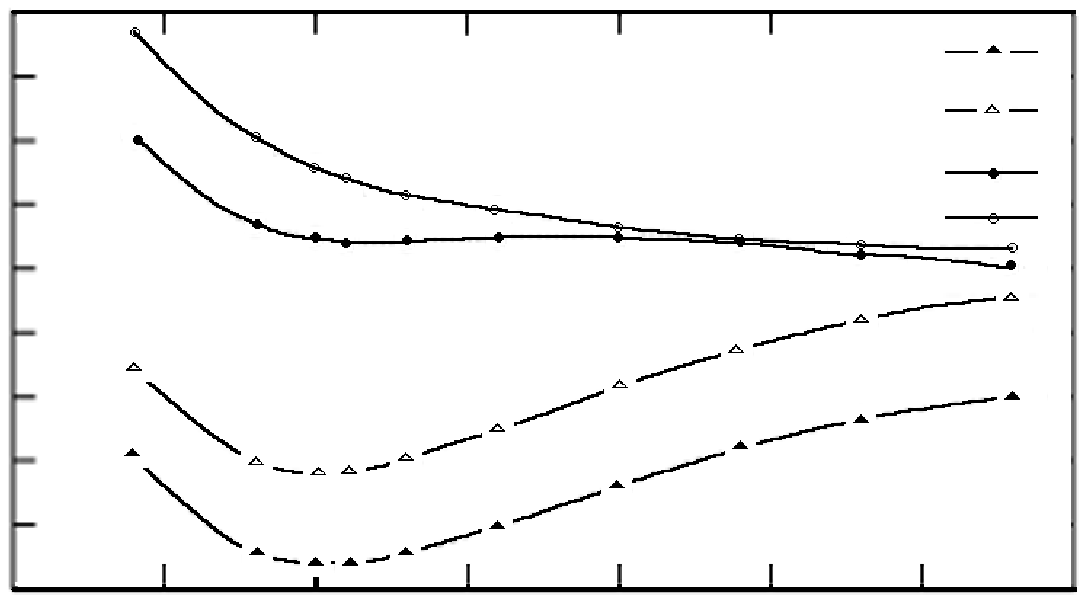}
\put(3094,1307){\makebox(0,0)[r]{$^2\Sigma_{1/2}^-$}}%
\put(3094,1457){\makebox(0,0)[r]{$^4\Sigma_{1/2}^-$}}%
\put(3094,1597){\makebox(0,0)[r]{$^2\Pi_{3/2}$}}%
\put(3094,1747){\makebox(0,0)[r]{$^2\Pi_{1/2}$}}%
\put(1925,2160){\makebox(0,0){The potential curves 
from GRECP/5e-MRD-CI calculations of PbH.}}%
\put(1925,2010){\makebox(0,0){Internuclear distance (X-axis) and total 
valence energy (Y-axis) are in a.u.}}%
\put(3450,100){\makebox(0,0){6}}%
\put(3014,100){\makebox(0,0){5.5}}%
\put(2579,100){\makebox(0,0){5}}%
\put(2143,100){\makebox(0,0){4.5}}%
\put(1707,100){\makebox(0,0){4}}%
\put(1271,100){\makebox(0,0){3.5}}%
\put(836,100){\makebox(0,0){3}}%
\put(400,100){\makebox(0,0){2.5}}%
\put(350,1860){\makebox(0,0)[r]{-3.82}}%
\put(350,1676){\makebox(0,0)[r]{-3.84}}%
\put(350,1491){\makebox(0,0)[r]{-3.86}}%
\put(350,1307){\makebox(0,0)[r]{-3.88}}%
\put(350,1122){\makebox(0,0)[r]{-3.9}}%
\put(350,938){\makebox(0,0)[r]{-3.92}}%
\put(350,753){\makebox(0,0)[r]{-3.94}}%
\put(350,569){\makebox(0,0)[r]{-3.96}}%
\put(350,384){\makebox(0,0)[r]{-3.98}}%
\put(350,200){\makebox(0,0)[r]{-4}}%
\end{picture}%
\endgroup

%% file: p2310tav.bbl
\begin{thebibliography}{99}
\bibitem{CNFSORT} The CNFSORT code for automatic selection of reference
		  configurations after the previous run for a group of states
		  ($n$) with $\sum_I C_{nI}^2=C^2_{n\Sigma}\approx
		  C^2_{Thr}$ (where $C^2_{Thr}$ is a given value)
		  is developed by E.Shulgina and A.Titov
		  employing the algorithm proposed by N.Mosyagin.
\end{thebibliography}
